\documentclass[letterpaper,titlepage,11pt]{article}
\usepackage{hyperref}
\usepackage{amssymb,amsmath,amsfonts}
\usepackage{epsfig}

\def\clock{{\count0=\time
           \divide\count0 60
           \ifnum\count0<10 0\fi\the\count0
           \multiply\count0 -60 \advance\count0 \time
           :\ifnum\count0<10 0\fi \the\count0
         }}
\newcommand{\timestamp}{{\small\vbox{\hbox{\tt\jobname.tex}
\hbox{\the\day/\the\month/\the\year, \clock}}}}

\newcommand{\be}{\begin{equation}} \newcommand{\ee}{\end{equation}}
\newcommand{\bea}{\begin{eqnarray}} \newcommand{\eea}{\end{eqnarray}}

\newcommand{\CO}{\mathcal{O}}
\newcommand{\CN}{\mathcal{N}}

\newcommand{\CT}{\mathcal{T}}

\newcommand{\CM}{\mathcal{M}}

\newcommand{\id}{\hbox{1\kern-.27em l}}
\newcommand{\sid}{\hbox{\scriptsize1\kern-.27em l}}

\newcommand{\we}{\kern-.1em\wedge\kern-.1em}
\newcommand{\scal}{\kern-.13em\cdot\kern-.13em}

\newcommand{\II}{I\kern-.09em I}

\newcommand{\R}{\mathbb{R}}
\newcommand{\T}{\mathbb{T}}

\newcommand{\spa}{\ , \ \ }

\newcommand{\Ord}{{\cal{O}}}

\newcommand{\mt}{\mathfrak{t}}
\newcommand{\ms}{\mathfrak{s}}

\newcommand{\hmt}{\hat{\mathfrak{t}}}
\newcommand{\hms}{\hat{\mathfrak{s}}}

\newcommand{\hmf}{\hat{\mathfrak{f}}}

\newcommand{\beastar}{\begin{eqnarray*}}
\newcommand{\eeastar}{\end{eqnarray*}}

\numberwithin{equation}{section}

\begin{document}

\begin{titlepage}

\rightline{\vbox{\small\hbox{\tt hep-th/0503021} }} \vskip 3cm

\centerline{\Large \bf New Phases of Thermal SYM and LST}
\vskip 0.2cm \centerline{\Large \bf from Kaluza-Klein Black Holes}

\vskip 1.6cm
\centerline{\bf Troels Harmark and Niels A. Obers}
\vskip 0.5cm
\centerline{\sl The Niels Bohr Institute}
\centerline{\sl Blegdamsvej 17, 2100 Copenhagen \O, Denmark}

\vskip 0.5cm

\centerline{\small\tt harmark@nbi.dk, obers@nbi.dk}

\vskip 1.6cm

\centerline{\bf Abstract} \vskip 0.2cm \noindent

We review the recently found map that takes any static and neutral
Kaluza-Klein black hole, i.e.
 any static and neutral black hole on Minkowski-space times a circle
 $\CM^d \times S^1$, and maps it to a  corresponding solution for
 a non- and near-extremal brane on a circle.
This gives a precise connection between phases of  Kaluza-Klein
black holes and the thermodynamic behavior of the
non-gravitational theories dual to  near-extremal branes on a circle. In
particular, for the thermodynamics of strongly-coupled
supersymmetric  Yang-Mills theories  on a circle we predict the
existence of a new non-uniform phase and find new information
about the localized phase. We also find evidence for the existence
of a new stable phase of $(2,0)$
 Little String Theory in the canonical ensemble
for temperatures above its Hagedorn temperature.


\end{titlepage}

\pagestyle{empty} \small
\normalsize

\pagestyle{plain} \setcounter{page}{1}


\section{Introduction}
It has been established in recent years that near-extremal
branes in string theory and M-theory provide
a link between black hole phenomena in gravity and the thermal physics
of non-gravitational theories.
Among the most prominent examples is
the duality between near-extremal D3-branes
and $\CN=4$ supersymmetric Yang-Mills theory
\cite{Maldacena:1997re}.
More generally, for all the supersymmetric branes in string/M-theory
one has dualities
between the near-extremal limit of the brane
and a non-gravitational theory \cite{Itzhaki:1998dd,Aharony:1999ti}.

In this talk we briefly review the recently found map \cite{Harmark:2004ws}
from static and neutral Kaluza-Klein black holes to near-extremal branes on a transverse
circle (see also Refs.~\cite{Bostock:2004mg,Aharony:2004ig} for related
work). This gives a precise connection between
phases of  Kaluza-Klein black holes and the
thermodynamic
behavior of certain non-gravitational theories.
The non-gravitational theories include
($p+1$)-dimensional supersymmetric Yang-Mills theories
with 16 supercharges compactified
on a circle, the uncompactified
($2+1$)-dimensional supersymmetric Yang-Mills theory
with 16 supercharges, and
$(2,0)$ Little String Theory.
\section{Review of phases of Kaluza-Klein black holes \label{secrev} }
We start with a short review of the current knowledge on phases of
static and neutral Kaluza-Klein black holes (see
\cite{Harmark:2004aa,Kol:2004ww} for recent reviews). These are
static solutions of
 the vacuum Einstein
equations (i.e. pure gravity) that have an event horizon, and
that asymptote to
Minkowski-space times a circle $\CM^d \times S^1$, i.e.
Kaluza-Klein space, with $d \geq 4$. The asymptotic behavior of these solutions
is characterized by the mass $M$ and the tension $\CT$ associated to the
compact direction \cite{Harmark:2003dg,Kol:2003if,Harmark:2004ch}.  Together with
the circumference $L$ of the circle at infinity, we can
construct two dimensionless parameters which we take as the
reduced mass $\mu = 16 \pi G_{\rm N}/L^{d-2}$ and relative tension
$n = L \CT/M$. The possible phases of Kaluza-Klein black holes can
then be drawn in a $(\mu,n)$ phase diagram, where all physically sensible
solutions fall in the range $\mu > 0$ and $ 0 \leq n \leq d-2$
\cite{Harmark:2003dg}.

The phase diagram appears to be divided in two separate regions
\cite{Elvang:2004iz,Harmark:2004aa}: (i) The region $0 \leq n \leq
1/(d-2)$, which contains solutions without Kaluza-Klein bubbles.
The solutions in this region have a local $SO(d-1)$ symmetry, and
hence two types of event horizon topologies, namely $S^{d-1}$ and
$S^{d-2} \times S^1$ for the black hole and
 string on a cylinder respectively. (ii) The region $1/(d-2) < n \leq d-2$ contains
solutions with Kaluza-Klein bubbles, which is the subject of
\cite{Elvang:2004iz}.

In this talk, we focus on the phases in region (i) and their mapping
to phases of non- and near-extremal branes. The mapping of the region
(ii) into phases of non- and near-extremal branes will be considered in
Ref.~\cite{Harmark:2004bb}.

At present, three branches are known for $0 \leq n \leq 1/(d-2)$: \newline
$\bullet$ {\it The uniform black string branch}. The metric is that of
a Schwarzschild black hole in $d$ dimensions times a compact $S^1$
direction. This branch has $n=1/(d-2)$ and exists for all $\mu$.
It is classically stable
for $\mu > \mu_{\rm GL}$ and classically unstable for $\mu < \mu_{\rm GL}$
where the Gregory-Laflamme mass $\mu_{\rm GL}$ can be obtained
numerically for each dimension $d$ \cite{Gregory:1993vy,Gregory:1994bj}.
See also \cite{Sorkin:2004qq,Kol:2004pn} for analysis of the
$d$-dependence of $\mu_{\rm GL}$.
\newline
$\bullet$ {\it The non-uniform black string branch}.
This branch was discovered in \cite{Gregory:1988nb,Gubser:2001ac}.
It starts at $\mu = \mu_{\rm GL}$ with $n=1/(d-2)$ in
the uniform string branch.
The approximate behavior near the Gregory-Laflamme point
$\mu=\mu_{\rm GL}$
is studied in \cite{Gubser:2001ac,Wiseman:2002zc,Sorkin:2004qq}.
For $4 \leq d \leq 9$ the results are that
the branch moves away from the Gregory-Laflamme point
according to
\begin{equation}
\label{nuni}
n (\mu) = \frac{1}{d-2} - \gamma (\mu - \mu_{\rm GL} )
+ \CO ((\mu - \mu_{\rm GL})^2 )
\spa 0 \leq  \mu -\mu_{\rm GL} \ll 1 \ ,
\end{equation}
where $\gamma > 0 $, so that the branch has decreasing $n$ and increasing
$\mu$. As shown in \cite{Harmark:2003dg} this means that the uniform
string branch has higher entropy than the non-uniform
string branch for a given mass.
(See for example the tables in \cite{Harmark:2004ws} for the numerical values of $\mu_{\rm GL}$ and
$\gamma$).
For $d=5$ a large piece of the branch was found
numerically in \cite{Wiseman:2002zc} thus providing detailed knowledge
of the behavior of the branch away from $\mu=\mu_{\rm GL}$.
\newline
$\bullet $ {\sl The black hole on cylinder branch.}
This branch has been studied analytically in
\cite{Harmark:2002tr,Harmark:2003yz,Kol:2003if,Harmark:2003eg,%
Gorbonos:2004uc} (see also \cite{Harmark:2003dg})
and numerically for $d=4$ in \cite{Sorkin:2003ka,Kudoh:2004hs}
and for $d=5$ in \cite{Kudoh:2003ki,Kudoh:2004hs}.
The branch starts in $(\mu,n)=(0,0)$ and then has increasing $n$ and $\mu$.
 In the limit of vanishing mass
(or, equivalently, very large $L$) this branch approaches the
solution of a Schwarzschild black hole in $d+1$ dimensions.
The first correction to the Schwarzschild black hole metric has
been found analytically in \cite{Harmark:2003yz,Gorbonos:2004uc}.
In particular, the branch starts off as \cite{Harmark:2003yz}
\begin{equation}
\label{nofM}
n (\mu)= \lambda_d
 \mu + \CO ( \mu^2 ) \spa \lambda_d =
  \frac{(d-2)\zeta(d-2)}{2(d-1)\Omega_{d-1}} \ .
\end{equation}
For $d=4$, the second order correction to the metric has also
been studied \cite{Karasik:2004ds}.
Moreover, recent numerical analysis \cite{Kudoh:2004hs} for $d=5$ shows
that the black hole branch meets the non-uniform black
string branch, indicating a topology changing transition
point \cite{Kol:2002xz,Kol:2003ja}.

Some further useful facts are the existence of a Smarr formula
\cite{Harmark:2003dg,Kol:2003if}, $\mt \ms = \frac{d-2-n}{d-1} \mu
$, where $\mt$, $\ms$ are the rescaled temperature $\mt = L T$ and
entropy $\ms = \frac{16 \pi G_{\rm N}}{L^{d-1}}$ respectively. In
particular, together with the first law of thermodynamics $\delta
\mu = \mt\, \delta \ms $, this means that, given a curve $n
(\mu)$, the entire thermodynamics can be obtained. We also note
that, as originally proposed in
\cite{Harmark:2002tr,Harmark:2003fz}, there exists a consistent
ansatz \cite{Wiseman:2002ti,Harmark:2003eg} that describes the
solutions with $0 \leq n \leq 1/(d-2)$. Finally we mention that
for any solution in this ansatz one can generate an infinite
number of copies \cite{Horowitz:2002dc,Harmark:2003eg}.
\section{Phases of non- and near-extremal branes}
Any Kaluza-Klein black hole in $d+1$ dimensions ($4 \leq d \leq
9$), can be mapped  \cite{Harmark:2004ws} to a corresponding brane
solution of Type IIA/B String Theory and M-theory, following the
method originally conceived in \cite{Hassan:1992mq}. In
particular, we obtain in this way a non-extremal $p$-brane on a
circle. These are thermal excitations of singly-charged extremal
1/2 BPS branes in String/M-theory with transverse space $\R^{d-1}
\times S^1$, i.e. with a transverse circle. A precise definition
of this class of branes, including a detailed discussion of the
physical parameters, thermodynamical relations and other
properties is given in Ref.~\cite{Harmark:2004ws}. These branes
are characterized by three independent dimensionless quantities,
the dimensionless mass $\bar {\mu}$, the relative tension
$\bar{n}$ and the charge $q$.

To obtain the map from Kaluza-Klein black holes on $\CM^d \times S^1$
to non-extremal $p$-branes on a circle one first uplifts the neutral
solution to 11 dimensions,  then performs a Lorentz boost and subsequently
uses a sequence of U-dualities.
The relation between $d$ and $p$ is $d + p +1 = D$ where $D=10$ or 11
for String theory or M-theory.
Since the focus in this talk is on near-extremal branes,
we refer to \cite{Harmark:2004ws} for the resulting expression of the
background obtained in this way as well as the associated map for the
physical quantities.

One can apply the map in particular to the ansatz describing the
class of Kaluza-Klein black holes with a local $SO(d-1)$ symmetry.
If the neutral solution is a black hole (black string)
then the topology of the horizon of the non-extremal $p$-brane solution
is $\R^{p} \times S^{d-1}$ ($\R^{p} \times S^{d-2} \times S^1$).
For this class of solutions the map was already discovered in
Ref.~\cite{Harmark:2002tr} at the level of equations of motion.
The boost/U-duality derivation thus provides us with a physical understanding of this
correspondence. Some of these maps have also been examined in
Refs.~\cite{Bostock:2004mg,Aharony:2004ig}.

We now wish to take the near-extremal limit of any non-extremal brane on a
circle, which gives us a corresponding near-extremal brane on a circle.
This is a non-extremal brane with infinitesimally
small temperature, or, equivalently, a non-extremal brane with
infinitely high charge. More precisely, we want to take a near-extremal
limit such that the size of the circle has the same scale as the
excitations of the energy above extremality.
This is because we want to keep the non-trivial physics related
to the presence of the circle.

For a non-extremal $p$-brane with volume $V_p$, circumference
$L$ and rescaled charge $q$, the near-extremal limit is then \cite{Harmark:2004ws}
\begin{equation}
\label{nelimit}
q \rightarrow \infty \spa L \rightarrow 0 \spa g
\equiv \frac{16\pi G_D}{V_p L^{d-2}}\ \, \mbox{fixed } \spa l
\equiv L \sqrt{q} \ \, \mbox{fixed } \spa
x^a\ \, \mbox{fixed } \ ,
\end{equation}
where $x_a$ are dimensionless transverse coordinates (obtained by
scaling with $L$). A near-extremal brane on a circle asymptotes
at infinity to a non-flat background which is
the near-horizon limit of the solution of extremal
branes on a circle. The latter is of course exactly known
using the superposition principle of BPS branes.
This space is also the one that we use  as the reference space
when computing the physical quantities, which are the energy \cite{Hawking:1996fd}
$E$ above
extremality  and the tension \cite{Harmark:2004ch} $\hat{\CT}$ in the circle direction.

In this way we have, in analogy with the $(\mu,n)$ phase diagram
of neutral Kaluza-Klein black holes, a two-dimensional phase diagram
for near-extremal branes. The quantities we use are the
rescaled energy $\epsilon = g E$ and tension
$ r = 2 \pi \hat{\CT}/E $, in terms of which we have
the $(\epsilon,r)$ phase diagram.
It is also useful to define the dimensionless versions of
the temperature $\hat{\mt} = l \, \hat{T}$ and entropy
$\hat{\ms} = \frac{g}{l} \hat{S}$
since $l$ and $g$ in \eqref{nelimit} have dimension length.
In terms of these the near-extremal Smarr formula takes the form
$\hat{\mt} \hat{\ms} = 2 \frac{d-2-r}{d} \epsilon $ and the
first law of thermodynamics is  $\delta \epsilon = \hmt \, \delta \hms$.
As a consequence, knowing a curve $r(\epsilon)$ in the near-extremal phase
diagram determines the entire thermodynamics.

We can apply now the near-extremal limit to the non-extremal
branes obtained via boost and U-duality from  neutral Kaluza-Klein black
holes. The resulting background is of the form \cite{Harmark:2004ws}
\begin{equation}
\label{gensolnh}
ds^2 = \hat H^{-\frac{d-2}{D-2}} \left( - U
dt^2 + \sum_{i=1}^p (du^i)^2 + \hat H V_{ab} d x^a d x^b
\right) \ ,
\end{equation}
\begin{equation}
\label{gensolnh2}
e^{2\phi} = \hat H^a \spa
A_{(p+1)} = \hat H^{-1}  dt \wedge du^1 \wedge \cdots \wedge du^p \ ,
\end{equation}
where $\hat H \propto 1-U$ and
$U$ and $V_{ab}$ are functions determining the metric of
the Kaluza-Klein black hole one started with.
The corresponding map from  the $(\mu,n)$ phase diagram of Kaluza-Klein black
holes to the $(\epsilon,r)$ phase diagram of near-extremal branes on a circle,
then takes the simple form
\begin{equation}
\label{nemap}
\epsilon = \frac{d+n}{2(d-1)} \mu \spa
r = 2 \frac{(d-1)n}{d+n}
\spa
\hat{\mt} =  \mt \sqrt{\mt \ms}
\spa
\hat{\ms} = \frac{\ms}{\sqrt{ \mt \ms}} \ .
\end{equation}
We also note the expression
\begin{equation}
\label{pf}
\mathfrak{p} = -\hmf  = \frac{d -4-2r}{d} \epsilon
\end{equation}
for the dimensionless pressure
in the world-volume directions of the near-extremal brane.
Here $\hmf$ is the rescaled free energy. In order to make
physically sense from the dual field theory point of view, the
pressure should be positive. Using \eqref{pf} this implies the interesting
bound $r \leq (d-4)/2$ or, equivalently, $n  \leq (d-4)/3$, which is
further examined in \cite{Harmark:2004ws}.

The three phases of Kaluza-Klein black holes reviewed in Section
\ref{secrev} thus immediately imply the existence of three corresponding
phases of non- and near-extremal branes on a circle: \newline
$\bullet$ {\it Uniform phase}. This is obtained by applying the map to the
uniform black string branch, thereby generating a non- or near-extremal $p$-brane
smeared on a circle. This solution, which is really a $(p+1)$-brane
since it is uniformly distributed on the transverse circle,
 is of course well known. \newline
$\bullet$ {\it Non-uniform phase}. By applying the map to the non-uniform black
string branch we find a new branch of solutions
for non- and near-extremal branes on a circle that are non-uniformly
distributed on the circle.
The physics of the neutral non-uniform string branch
near $\mu = \mu_{\rm GL}$
is captured by the formula \eqref{nuni}.
Using the map from the neutral case
to the non-extremal case one finds that the new non-extremal branch
emerges out of the uniform phase (with $\bar{n} = 1/(d-2)$)
at some critical mass $\bar{\mu}_{\rm c} (\mu,q)$, which has
the natural interpretation as being a Gregory-Laflamme critical mass of
non-extremal branes uniformly distributed on a circle.

For the near-extremal case we find likewise from \eqref{nuni} and
\eqref{nemap} that the Gregory-Laflamme point is mapped to the
point $(\epsilon_c,2/(d-1))$ in the $(\epsilon,r)$
phase diagram with critical energy $\epsilon_c = \frac{d-1}{2(d-2)} \mu_{\rm GL}$.
Moreover, the first part of the non-uniform near-extremal branch is
described by
\begin{equation}
\label{nearr}
r (\epsilon) = \frac{2}{d-1} - \hat{\gamma} ( \epsilon - \epsilon_c )
+ \CO ( ( \epsilon - \epsilon_c )^2 )
\spa 0 \leq \epsilon-\epsilon_c \ll 1 \ ,
\end{equation}
with $\hat{\gamma}$ a $d$-dependent constant determined by $\gamma$ and $\mu_{\rm GL}$
>From \eqref{nearr} all other physical information, such as the entropy
and free energy near the critical point can be derived \cite{Harmark:2004ws}.

The existence of this new non-uniform phase suggests that
near-extremal branes smeared on a circle have a critical energy below which
they are classically unstable (see also Refs.~\cite{Bostock:2004mg,Aharony:2004ig}).
The specific heat of the uniform
near-extremal branch is, however, positive. It would be interesting
to study this further in connection with the Gubser-Mitra conjecture%
\footnote{For non-extremal branes that are near extremality (without
the near-extremal limit taken) the Gubser-Mitra conjecture is satisfied since,
although the specific heat is positive, the quantity
$(\partial \nu /\partial Q )_T$ is negative and hence the brane is
not thermodynamically stable.}
\cite{Gubser:2000ec,Gubser:2000mm,Reall:2001ag,Gregory:2001bd}
(see also the recent paper \cite{Gubser:2004dr}). \newline
$\bullet$ {\it Localized phase}. Finally, we can apply the map to the
neutral black hole branch for which the first correction to the metric
and the thermodynamics is known, generating non- and near-extremal
$p$-branes localized on a circle. This phase is not new, as it is expected on
general physical grounds. However, we are now able to compute the
first correction to the metric and thermodynamics of this background.

For a given charge $q$, the branch of non-extremal branes localized on a
circle starts in the  extremal point $(\bar{\mu},\bar{n}) = (q,0)$ and
goes up with a slope $2(d-1)\lambda_d/d$. Thus, a non-extremal
$p$-brane localized on a circle becomes point-like in the extremal
limit, i.e. for small temperatures.

For the near-extremal case we find from \eqref{nofM} and
\eqref{nemap} the leading behavior of the localized branch as
\begin{equation}
\label{rofeps}
r (\epsilon) = \frac{4(d-1)^2}{d^2} \lambda_d \epsilon + \Ord (\epsilon^2) \ ,
\end{equation}
from which the entropy and free energy can be computed \cite{Harmark:2004ws}.

For the particular case
of the M5-brane on a circle we can go even further and use
the numerically obtained $d=5$ non-uniform  branch of Wiseman
\cite{Wiseman:2002zc} to obtain the
corresponding near-extremal non-uniform phase. Similarly, the
new numerical results of \cite{Kudoh:2004hs} can be used to also determine
the complete near-extremal localized phase. This is currently in progress.
\section{Phases of thermal SYM and LST}
The results for the non-uniform and localized phase
 of near-extremal branes are particularly interesting,
since they provide us with new information about the
dual non-gravitational theories at finite temperature%
\footnote{See also \cite{Susskind:1997dr,Barbon:1998cr,Li:1998jy,%
Martinec:1998ja,Martinec:1999bf,Fidkowski:2004fc,Aharony:2004ig} and
references therein.}.
In particular, we have studied \cite{Harmark:2004ws}: \newline
$\blacktriangleleft$ The M5-brane on a circle, which is dual to thermal (2,0)
Little String Theory (LST). \newline
$\blacktriangleleft$ The D$(p-1)$-brane
on circle, which is dual to thermal $(p+1)$-dimensional
supersymmetric Yang-Mills (SYM) theory on $\R^{p-1} \times S^1$. \newline
$\blacktriangleleft$ The M2-brane on a circle, which is dual to
thermal $(2+1)$-dimensional SYM theory on $\R^2$.

In these dual non-gravitational theories,
the localized phase corresponds to the low temperature/low
energy regime of the dual theory, whereas the uniform phase corresponds
to the high temperature/high energy regime of the theory.
The non-uniform phase
appears instead for intermediate temperatures/energies.

In particular, by translating the results for the thermodynamics
of the localized and non-uniform phase in terms of the dual
non-gravitational theories we find: \newline
$\bullet $ The first correction to the thermodynamics for the localized
phase of the SYM theories.
The dimensionless expansion parameter in
the localized phase is $\hat T/\hat T_0$ with
$ \hat T_0 =\sqrt{2\pi} (\lambda \hat L^{3-p} )^{-1/2} \hat L^{-1}$,
with $\lambda$ the 't
Hooft coupling of the gauge theory and $\hat L$ the circumference of the
field theory circle. \newline
$\bullet$ A prediction of a new non-uniform phase of the SYM theories
including the first correction around the point where the non-uniform
phase emanates from the uniform phase.
The critical temperature that characterizes
the emergence of
 the non-uniform phase is
$\hat T_c =  \hat T_0 \hmt_c$, with $\hmt_c$ a numerically
determined constant that depends on $p$. In Ref.~\cite{Aharony:2004ig} evidence
for the existence of this non-uniform phase at weak 't Hooft coupling
was presented by considering (1+1)-dimensional SYM on $\T^2$. Moreover, it
was shown that the critical
temperature behaves as $\hat{T}_c \sim 1/\lambda$ as opposed to the predicted
strong coupling behavior $\hat{T}_c \sim 1/\sqrt{\lambda}$.
\newline
$\bullet $ Using the numerical data of Wiseman \cite{Wiseman:2002zc}
for the $d=5$ non-uniform branch we have numerically
 computed the corresponding thermodynamics in
$(2,0)$ LST. This gives
 a new stable phase of $(2,0)$ LST in the canonical ensemble,
for temperatures above its Hagedorn temperature $\hat T_c = T_{\rm hg}$.
We furthermore computed the first correction to the
thermodynamics in the infrared region, when moving away from
the infrared fixed point, which is superconformal $(2,0)$
theory.


{\bf Acknowledgement}

 TH would like to thank the organizers of the RTN Workshop in Kolymbari,
Crete, September 5-10, 2004, for the opportunity to present this
work. NO would like to thank the organizers of the 37th
International Symposium Ahrenshoop, Wernsdorf, August 23-27, 2004,
for the stimulating atmosphere and the invitation to present this
work. Work partially supported by the European Community's Human
Potential Programme under contract MRTN-CT-2004-005104
`Constituents, fundamental forces and symmetries of the universe'.


\providecommand{\href}[2]{#2}\begingroup\raggedright\endgroup

\end{document}